\documentclass[paper,notoc]{JHEP3}
\usepackage{amssymb}   
\usepackage{amsmath}
\usepackage{graphicx}

\newcommand{\ord}{\begin{cal}O\end{cal}}

\newcommand{\eps}{\epsilon}

\def\beq{\begin{equation}}
\def\eeq{\end{equation}}
\def\bsp#1\esp{\begin{split}#1\end{split}}

\newcommand{\pasb}[1]{\left(\frac{\alpha_s}{\pi}\right)^{#1}}
\newcommand{\Dplus}[1]{\left[\frac{\log^{#1}(1-z)}{1-z}\right]_+}
\newcommand{\DplusOne}[0]{\left[\frac{\log(1-z)}{1-z}\right]_+}
\newcommand{\DplusZero}[0]{\left[\frac{1}{1-z}\right]_+}

\title{
Higgs boson gluon-fusion production beyond threshold in N$^3$LO QCD
}

\author{Charalampos Anastasiou$^a$, Claude Duhr$^b$, Falko Dulat$^a$, Elisabetta Furlan$^c$, Thomas Gehrmann$^d$, 
Franz Herzog$^e$, Bernhard Mistlberger$^a$\\
{}$^a$Institute for Theoretical Physics, ETH Z\"urich,
  8093 Z\"urich, Switzerland 
\\
{}$^b$Center for Cosmology, Particle Physics and Phenomenology (CP3),\\
\phantom{{}$^b$}Universit\'e catholique de Louvain,\\  
\phantom{{}$^b$}Chemin du Cyclotron 2, 1348 Louvain-La-Neuve, Belgium
\\
{}$^c$Fermilab, Batavia, IL 60510, USA   
\\
{}$^d$Physik-Institut, Universit\"at Z\"urich, 
Winterthurerstrasse 190, 8057 Z\"urich, Switzerland 
{}$^e$Nikhef, Science Park 105, NL-1098 XG Amsterdam, The Netherlands\\
\phantom{{}$^e$}CERN Theory Division, CH-1211, Geneva 23, Switzerland
}

\preprint{CP3-14-71, ZU-TH 39/14, FERMILAB-PUB-14-461-T, NIKHEF 2014-048, CERN-PH-TH-2014-221}

\abstract{ 
In this article, we compute the gluon fusion Higgs boson cross-section at 
N$^3$LO through the second term in the threshold expansion.   
This calculation constitutes a major milestone towards the full N$^3$LO cross section.
Our result has the best formal accuracy in the threshold expansion currently available, and includes contributions from collinear regions besides subleading corrections from soft and hard regions, as well as certain logarithmically enhanced contributions for general kinematics. We use our results to perform a critical appraisal of the validity of the threshold approximation at N$^3$LO in perturbative QCD.
}


\begin{document}

\section{Introduction}
\label{sec:introduction}

With the discovery of the Higgs boson~\cite{HiggsDiscovery}, the 
Standard Model is a fully predictive theory, with all of its
parameters determined experimentally.    
This fact renders the total Higgs boson production 
cross-section an excellent precision test of the theory. Theoretical 
predictions for the inclusive cross-section therefore play an important role in
measurements of Higgs-boson observables in general and in the determination of the 
coupling strengths of the Higgs boson in particular. 

For this reason,
obtaining a reliable theoretical estimate of the gluon-fusion cross-section, 
the dominant production mechanism of a Higgs boson at the LHC, has been a major objective in
perturbative QCD for the last decades. The
very large size of  the next-to-leading-order (NLO) perturbative corrections in the strong 
coupling $\alpha_s$ indicated a slow convergence of the  $\alpha_s$ 
expansion~\cite{nlo}. 
The smaller size of the next-to-next-to-leading order (NNLO) corrections
inspired some confidence that QCD effects beyond NNLO may be smaller 
than $\pm 10\%$, as indicated from the variation of the renormalization
 scale~\cite{nnlo}. On the basis of this belief,  further refinements of
 the cross-section with electroweak corrections and finite quark-mass
 effects  (at a  $\sim 5\%$ level of precision) followed~\cite{ewktopbottom}. 

Currently, no full computation of the hadronic Higgs-boson cross-section is available at next-to-next-to-next-to-leading order N$^3$LO.
It is possible to obtain some information on the missing higher orders beyond NNLO in the so-called 
threshold limit where the Higgs boson is predominantly produced at threshold
and the additional QCD radiation is soft. In this limit, soft QCD emissions factorize from the hard interaction
and can be resummed~\cite{nnloresummation}.  
After the completion of the NNLO corrections~\cite{nnlo}, it was observed that
the threshold approximation can be made to capture the bulk of the perturbative corrections through NNLO. 
It is then tantalising to speculate 
if a similar approximation is sufficient to predict the value of the Higgs-boson cross-section at N$^3$LO in QCD. 
Recently, various approximate N$^3$LO cross-section estimates were put forward which rely crucially on the threshold assumption~\cite{Moch:2005ky,Ball:2013bra,deFlorian:2014vta}. Given these considerations, it is important to quantify the reliability of the threshold approximation at N$^3$LO.

Logarithmically enhanced threshold contributions at N$^3$LO to the cross-section coming from the emission of
soft gluons have been computed almost a decade ago~\cite{Moch:2005ky}.
A few months ago, we completed the computation of the first term in the threshold expansion,
the so-called soft-virtual term,
by computing in addition the constant term proportional to $\delta(1-z)$~\cite{Anastasiou:2014vaa}, 
which includes in particular the complete 
three-loop corrections to Higgs production via gluon fusion~\cite{formfactor}.
Recently, some further logarithmic
corrections which belong to the second order in the threshold
expansion were conjectured in ref.~\cite{deFlorian:2014vta,vogtqg}.  
In this paper we compute for the first time the complete second order in the threshold 
expansion. 
This result is an important step in the direction of the computation of the 
N$^3$LO cross-section for arbitrary values of $z$, a goal which has only been achieved so far 
at N$^3$LO for the three-loop corrections and the single-emission contributions at two loops~\cite{Anastasiou:2013mca,RVV_bf,RVV_ct}.
We combine the knowledge of the single-real emission contributions with the ultra-violet and parton-density counterterms to obtain the exact result for the first three logarithmically-enhanced terms beyond the 
soft-virtual approximation. Both results combined are not only a major milestone towards the complete Higgs-boson cross-section at N$^3$LO, but they also constitute the most precise
calculation of the Higgs-boson cross-section at N$^3$LO beyond threshold.

In a second part of our paper, we use our results and perform a critical appraisal of the threshold approximation. We define a way to quantify the convergence of the truncated threshold expansion, and we perform a numerical study of the convergence of the threshold expansion at NLO, NNLO and N$^3$LO. Given the widely accepted dominance of the threshold limit in Higgs production at the LHC, our study is an important ingredient to asses the reliability of the threshold approximation at N$^3$LO in QCD.

This paper is organised as follows: In Section~\ref{sec:results} we present our results for the complete second term in the threshold expansion and the exact results for the coefficients of the first three leading logarithmically-enhanced terms in the threshold limit. In Section~\ref{sec:numerics} we perform a critical appraisal of the threshold expansion, both in $z$-space and in Mellin-space. In Section~\ref{sec:conclusions} we draw our conclusions.

\section{ Analytic results for the N$^3$LO partonic cross-section} 
\label{sec:results}
\subsection{The gluon-fusion cross-section}
In this section we present the main results of our paper. We start by giving a short review of the inclusive gluon-fusion cross-section and its analytic properties, and then we present our results in subsequent sections.

The inclusive cross-section $\sigma$ for the production of a Higgs boson is given by  
\begin{equation}
\label{eq:sigma}
\sigma = \tau \sum_{ij} \left( f_i \otimes f_j  \otimes 
\frac{\hat{\sigma}_{ij}(z)}{z}
\right)\left(\tau \right) \,,
\end{equation} 
where  $\hat{\sigma}_{ij}$ are the partonic cross-sections for
producing a Higgs boson from the parton species $i$ and $j$, and $f_i$ and
$f_j$ are the corresponding parton densities. We have defined the ratios
\beq
\tau = \frac{m_H^2}{S} {\rm~~and~~} z = \frac{m_H^2}{s}\,,
\eeq
where $m_H$ denotes the Higgs-boson mass and $s$ and $S$ denote the squared partonic and hadronic center-of-mass energies.
 The convolution of two
functions is  defined as
\begin{equation}\label{eq:convolution}
(A \otimes B)(\tau) = \int_0^1 dx \,dy\, A(x)\,B(y)\, \delta(\tau - x y)\,.
\end{equation}
In the rest of this section we only concentrate on the partonic cross-sections. If we work in perturbative QCD, and after integrating out the top quark, the 
partonic cross-sections take the form
\beq
\label{eq:sigma_partonic}
\frac{\hat{\sigma}_{ij}(z)}{z} = \frac{\pi\,C^2}{8V}\,
\sum_{k=0}^\infty\pasb{k}\,\eta_{ij}^{(k)}(z)
\,,
\eeq
with $V=N_c^2-1$ and $N_c$ the number of $SU(N_c)$ colours, and
 $C\equiv C(\mu^2)$ and $\alpha_s\equiv \alpha_s(\mu^2)$ denote the Wilson coefficient~\cite{Wilson} and the strong coupling constant, evaluated at the scale $\mu^2$. At leading order in $\alpha_s$ only the 
gluon-gluon initial state contributes,
$\eta_{ij}^{(0)}(z) = \delta_{ig}\,\delta_{jg}\,\delta(1-z)$.
The partonic cross-sections through NNLO, $\eta_{ij}^{(1,2)}(z)$, 
can be found in ref.~\cite{nnlo}. 

Before presenting our results, let us discuss some general properties of the N$^3$LO coefficients $\eta_{ij}^{(3)}(z)$ which will be useful in the remainder of this section. First, $\eta_{ij}^{(3)}(z)$ does not only contain the three-loop corrections to inclusive Higgs production, but also contributions from the emission of up to three partons in the final state at the same order in perturbation theory. So far, only the single-emission contributions at two loops are known for generic values of $z$~\cite{Anastasiou:2013mca,RVV_bf,RVV_ct,Gehrmann:2011aa,Duhr:2013msa}, and only a few terms in the threshold expansion for  the contributions with up to two additional partons in the final state are known~\cite{Anastasiou:2014vaa,triplereal,Li:2014bfa}. Each of these contributions is ultra-violet (UV) and infra-red (IR) divergent, and the divergences manifest themself as poles in the dimensional regulator $\eps$.

While the first three leading poles at N$^3$LO cancel when summing over all the contributions, the coefficient of $\eps^{-3}$ is non-zero. These remaining divergences cancel when suitable UV and IR counterterms are included. We generically write
\beq
\label{eq:etadec}
\eta^{(3)}_{ij}(z)=\Delta_{ij}^{(3)}(z,\epsilon)+\chi_{ij}^{(3)}(z,\epsilon)\,,
\eeq
where $\Delta_{ij}^{(3)}(z,\epsilon)$ is the combined UV and IR counterterm and $\chi_{ij}^{(3)}(z,\epsilon)$ is the (bare) contribution from the different particle multiplicities at N$^3$LO. Note that each term in the right-hand side has poles at $\eps=0$, but the sum is finite. The counterterm is determined completely from lower orders\footnote{There is a typo in eq.~(2.8) of ref.~\cite{Buehler:2013fha}. The combination $3\,P^{(0)}_{ik}\otimes P^{(1)}_{kj}+3\,P^{(1)}_{ik}\otimes P^{(0)}_{kj}$ in the fourth line should be replaced by $2\,P^{(0)}_{ik}\otimes P^{(1)}_{kj}+4\,P^{(1)}_{ik}\otimes P^{(0)}_{kj}$.}~\cite{NNLOXsec,Buehler:2013fha}, as well as the QCD $\beta$ function~\cite{UV} and the three-loop splitting functions~\cite{Moch:2004pa}.

The contributions arising from different multiplicities can be separated into six different terms as 
\beq\label{eq:chi_exp}
\chi_{ij}^{(3)}(z,\epsilon)=\chi^{(3,0)}_{ij}(\epsilon)\,\delta(1-z)+\sum\limits_{m=2}^6 (1-z)^{-m \epsilon}\,\chi^{(3,m)}_{ij}(z,\epsilon).
\eeq
where the functions $\chi^{(3,m)}_{ij}(z,\epsilon)$ are meromorphic  with at most a simple pole at $z=1$.
While the first term only contributes at threshold and contains the entirety of the three-loop corrections, the second term receives 
contributions from all additional parton emissions.

The partonic cross-sections are convoluted with the parton luminosities, and the pole at $z=1$ in the gluon-gluon initial state introduces a divergence in the integrand as $z\rightarrow 1$. 
The singularities are regulated in dimensional regularisation by expanding the factors $(1-z)^{-1-m\epsilon}$ in terms of delta functions and plus-distributions.
\beq
\label{eq:plusd}
(1-z)^{-1-m\epsilon}=-\frac{1}{m\epsilon}\delta(1-z)+\sum\limits_{j=0}^\infty\frac{(-m\epsilon)^j}{j!}\left[\frac{\log^j(1-z)}{1-z}\right]_+\,,
\eeq
where the plus-distribution is defined by its action on a test function $\phi(z)$.
\beq
\int\limits_0^1 dz \left[\frac{\log^j(1-z)}{1-z}\right]_+ \phi (z)\equiv\int \limits_0^1 dz \frac{\log^j(1-z)}{1-z}\left[\phi(z)-\phi(1)\right].
\eeq
In order to expose the distributions, we write
\beq\label{eq:singsplit}
\chi_{ij}^{(3,m)}(z,\eps) = \chi_{ij}^{(3,m),\textrm{sing}}(z,\eps) + \chi_{ij}^{(3,m),\textrm{reg}}(z,\eps)\,,
\eeq 
with $\chi_{ij}^{(3,m),\textrm{sing}}(z,\eps)$ the residue at $z=1$ (divided by $(z-1)$). The singular contribution is only non-zero for the gluon-gluon initial state.

Similar to eq.~\eqref{eq:singsplit}, we can split the partonic cross-sections into a singular and a regular part,
\begin{equation}
\eta_{ij}^{(3)}(z)  = \eta_{ij}^{(3),\textrm{sing}}(z) + \eta_{ij}^{(3),\textrm{reg}}(z)\,,
\end{equation}
where the singular contribution is precisely the cross-section at threshold~\cite{Moch:2005ky,Anastasiou:2014vaa} and the regular term describes terms that are formally subleading 
and take the form of a polynomial in $\log(1-z)$,
\beq\label{eq:eta_reg_log}
\eta_{ij}^{(3),\textrm{reg}}(z) = \sum_{m=0}^5\log^m(1-z)\,\eta_{ij}^{(3,m),\textrm{reg}}(z) \,,
\eeq
where the $\eta_{ij}^{(3,m),\textrm{reg}}(z)$ are holomorphic in a neighbourhood of $z=1$.
The coefficients of these logarithms are the main subject of this paper, and in the rest of this section we show how to explicitly determine some of the regular coefficients of the threshold logarithms.

\subsection{Next-to-soft corrections}
All the regular terms are formally subleading in the threshold expansion compared to the soft-virtual term. If we want to compute these subleading corrections, we need to know the counterterms $\Delta_{ij}^{(3)}(z,\eps)$ and all process with different multiplicities contributing to $\chi_{ij}^{(3)}(z,\eps)$. To date, however, only the counterterms and the single-emission contributions are known for arbitrary values of $z$.  Since the coefficients of the logarithms are holomorphic, they admit a Taylor expansion around $z=1$. In this section we discuss how to approximate the coefficients of the logarithms by their threshold expansion around $z=1$. In particular, one of the main results of this paper is the complete computation of the first subleading term in the threshold expansion, corresponding to the value at $z=1$ of the coefficients in eq.~\eqref{eq:eta_reg_log} and dubbed the \emph{next-to-soft} term in the remainder of this paper. Note that the next-to-soft term receives 
for the first time contributions from the quark-gluon (and anti-quark-gluon) initial state besides the gluon-gluon initial state.

In ref.~\cite{Anastasiou:2014vaa} the next-to-soft term of the triple-emission contribution was computed. Hence, we are only missing the next-to-soft corrections to the double-emission contribution at one-loop. We have recently completed the computation of all the relevant diagrams contributing to the next-to-soft term. In the following we only present the results of the computation, and details of the computation will be given elsewhere. Here it suffices to say that, unlike the contribution to the soft-virtual term~\cite{Anastasiou:2014vaa,Li:2014bfa}, we also need to consider contributions from regions where the virtual gluon can be collinear to one of the external partons besides subleading corrections to the soft and hard regions. In the following we present the next-to-soft cross-sections 
$\left. \eta_{ij}^{(3)}(z) \right|_{(1-z)^{0}}$ for values of the  renormalization and factorization scales equal to the Higgs mass. 
The corresponding expressions for arbitrary scales can be derived easily from renormalization group and DGLAP evolution.  
We find
\begin{align}\label{eq:ggN3LONSoft}
&\eta_{gg}^{(3)}(z)_{\big|(1-z)^0} = -8\,N_c^3\,\log^5(1-z)
+\left(\frac{353}{9}N_c^3-\frac{20}{9}N_c^2N_f\right)\,\log^4(1-z)\\
&\qquad+\Bigg[ \left(56\,\zeta_2-\frac{3469}{54}\right)N_c^3 + \frac{205}{18}N_c^2N_f-\frac{4}{27}N_cN_f^2\Bigg]\,\,\log^3(1-z)\nonumber\\
&\qquad
+\Bigg\{\left(-181\,\zeta_3-\frac{2147}{12}\,\zeta_2+\frac{2711}{27}\right)N_c^3
+\left[\left(\frac{545}{48}\,\zeta_2-\frac{4139}{216}\right)N_c^2+\frac{1}{4}\right]N_f\nonumber\\
&\qquad\quad+\frac{59}{108}N_cN_f^2\Bigg\}\,\log^2(1-z)\nonumber\\
&\qquad
+\Bigg\{
\left(77\,\zeta_4+362\,\zeta_3+\frac{2375}{18}\,\zeta_2-\frac{9547}{108}\right)N_c^3
+\left[\left(-\frac{223}{12}\,\zeta_3-\frac{1813}{72}\,\zeta_2+\frac{8071}{324}\right)N_c^2\right.\nonumber\\
&\qquad\quad\left.+3\,\zeta_3+\frac{1}{24}\,\zeta_2-\frac{17}{4}\right]N_f
+\left(\frac{4}{9}\,\zeta_2-\frac{163}{324}\right)N_cN_f^2\Bigg\}\,\log(1-z) \nonumber\\
&\qquad+\left(-186\,\zeta_5+\frac{725}{6}\,\zeta_2\,\zeta_3-\frac{821}{12}\,\zeta_4-\frac{32849}{216}\,\zeta_3-\frac{11183}{162}\,\zeta_2+\frac{834419}{23328}\right)N_c^3\nonumber\\
&\qquad\quad+\left[\left(\frac{19}{8}\,\zeta_4+\frac{1789}{72}\,\zeta_3+\frac{4579}{324}\,\zeta_2-\frac{527831}{46656}\right)N_c^2-\frac{1}{4}\,\zeta_4-\frac{149}{72}\,\zeta_3-\frac{5}{24}\,\zeta_2+\frac{5065}{1728}\right]N_f\nonumber\\
&\qquad\quad+\left(-\frac{5}{27}\,\zeta_3-\frac{19}{36}\,\zeta_2+\frac{49}{729}\right)N_cN_f^2
\,.\nonumber
\end{align}
\allowdisplaybreaks
\begin{align}\label{eq:qgN3LONSoft}
&\eta_{qg}^{(3)}(z)_{\big|(1-z)^0}= \Bigg(\frac{587N_c^3}{768}-\frac{247N_c}{256}+\frac{181}{768N_c}-\frac{9}{256N_c^3}\Bigg)\,\log^5(1-z)\\
&\qquad+\Bigg[-\frac{9155N_c^3}{27648}+\frac{899N_c}{1024}-\frac{15805}{27648N_c}+\frac{229}{9216N_c^3}\nonumber\\
&\qquad\quad+\left(\frac{803N_c^2}{6912}-\frac{11}{72}+\frac{253}{6912N_c^2}\right)N_f\Bigg]\,\log^4(1-z)\nonumber\\
&\qquad+\Bigg[ \left(-\frac{2791}{576}\,\zeta_2+\frac{166903}{41472}\right)N_c^3+\left(\frac{3839}{576}\,\zeta_2-\frac{57691}{13824}\right)N_c+\left(-\frac{1241}{576}\,\zeta_2+\frac{473}{41472}\right)\frac{1}{N_c}\nonumber\\
&\qquad\quad+\left(\frac{193}{576}\,\zeta_2+\frac{211}{1536}\right)\frac{1}{N_c^3}+\left(-\frac{1837N_c^2}{2592}+\frac{361}{432}-\frac{329}{2592N_c^2}\right)N_f\nonumber\\
&\qquad\quad+\left(\frac{7N_c}{864}-\frac{7}{864N_c}\right)N_f^2\Bigg]\,\,\log^3(1-z)\nonumber\\
&\qquad
+\Bigg\{\left(\frac{1687}{96}\,\zeta_3+\frac{1729}{576}\,\zeta_2-\frac{120073}{41472}\right)N_c^3+\left(-\frac{4241}{192}\,\zeta_3-\frac{1589}{288}\,\zeta_2+\frac{46025}{13824}\right)N_c\nonumber\\
&\qquad\quad+\left(\frac{485}{96}\,\zeta_3+\frac{541}{192}\,\zeta_2-\frac{14087}{41472}\right)\frac{1}{N_c}+\left(-\frac{103}{192}\,\zeta_3-\frac{29}{96}\,\zeta_2-\frac{145}{1536}\right)\frac{1}{N_c^3}\nonumber\\
&\qquad\quad
+\left[\left(-\frac{185}{288}\,\zeta_2+\frac{6427}{10368}\right)N_c^2+\frac{59}{72}\,\zeta_2-\frac{215}{288}+\left(-\frac{17}{96}\,\zeta_2+\frac{1313}{10368}\right)\frac{1}{N_c^2}\right]N_f\nonumber\\
&\qquad\quad+\left(-\frac{11N_c}{432}+\frac{11}{432N_c}\right)N_f^2\Bigg\}\,\log^2(1-z)\nonumber\\
&\qquad+\Bigg\{\left(-\frac{871}{96}\,\zeta_4-\frac{283}{72}\,\zeta_3-\frac{3755}{1152}\,\zeta_2+\frac{1641013}{248832}\right)N_c^3\nonumber\\
&\qquad\quad+\left(\frac{3787}{384}\,\zeta_4+\frac{2297}{288}\,\zeta_3+\frac{20545}{3456}\,\zeta_2-\frac{46859}{9216}\right)N_c\nonumber\\
&\qquad\quad+\left(-\frac{53}{96}\,\zeta_4-\frac{85}{18}\,\zeta_3-\frac{7039}{3456}\,\zeta_2-\frac{340909}{248832}\right)\frac{1}{N_c}\nonumber\\
&\qquad\quad+\left(-\frac{91}{384}\,\zeta_4+\frac{65}{96}\,\zeta_3-\frac{83}{128}\,\zeta_2-\frac{431}{3072}\right)\frac{1}{N_c^3}\nonumber\\
&\qquad\quad+\left[\left(\frac{125}{144}\,\zeta_3+\frac{155}{288}\,\zeta_2-\frac{157411}{62208}\right)N_c^2-\frac{55}{36}\,\zeta_3-\frac{473}{432}\,\zeta_2+\frac{9859}{3456}\right.\nonumber\\
&\qquad\quad\left.+\left(\frac{95}{144}\,\zeta_3+\frac{481}{864}\,\zeta_2-\frac{20051}{62208}\right)\frac{1}{N_c^2}\right]N_f+\left(\frac{29N_c}{432}-\frac{29}{432N_c}\right)N_f^2\Bigg\}\,\log(1-z) \nonumber\\
&\qquad +\left(\frac{1687}{96}\,\zeta_5-\frac{505}{48}\,\zeta_2\,\zeta_3-\frac{649}{2304}\,\zeta_4+\frac{34117}{3456}\,\zeta_3+\frac{3691}{1296}\,\zeta_2-\frac{1457441}{995328}\right)N_c^3\nonumber\\
&\qquad\quad+\left(-\frac{1447}{64}\,\zeta_5+\frac{2807}{192}\,\zeta_2\,\zeta_3-\frac{73}{64}\,\zeta_4-\frac{4001}{432}\,\zeta_3-\frac{5833}{1296}\,\zeta_2+\frac{53237}{995328}\right)N_c\nonumber\\
&\qquad\quad+\left(\frac{545}{96}\,\zeta_5-\frac{55}{12}\,\zeta_2\,\zeta_3+\frac{2245}{2304}\,\zeta_4-\frac{463}{1152}\,\zeta_3+\frac{95}{72}\,\zeta_2+\frac{422195}{331776}\right)\frac{1}{N_c}\nonumber\\
&\qquad\quad+\left(-\frac{41}{64}\,\zeta_5+\frac{31}{64}\,\zeta_2\,\zeta_3+\frac{43}{96}\,\zeta_4-\frac{5}{24}\,\zeta_3+\frac{1}{3}\,\zeta_2+\frac{1699}{12288}\right)\frac{1}{N_c^3}\nonumber\\
&\qquad\quad
+\left[\left(\frac{193}{576}\,\zeta_4-\frac{47}{27}\,\zeta_3-\frac{139}{324}\,\zeta_2+\frac{82171}{248832}\right)N_c^2-\frac{5}{32}\,\zeta_4+\frac{1723}{864}\,\zeta_3+\frac{229}{324}\,\zeta_2-\frac{17219}{124416}\right.\nonumber\\
&\qquad\quad\left.+\left(-\frac{103}{576}\,\zeta_4-\frac{73}{288}\,\zeta_3-\frac{5}{18}\,\zeta_2-\frac{15911}{82944}\right)\frac{1}{N_c^2}\right]N_f\nonumber\\
&\qquad\quad
+\left[\left(-\frac{1}{72}\,\zeta_3-\frac{125}{3888}\right)N_c+\left(\frac{1}{72}\,\zeta_3+\frac{125}{3888}\right)\frac{1}{N_c}\right]N_f^2\,.\nonumber
\end{align}

The leading logarithms in the above equations  can be compared with recent results in the literature. 
The coefficients of $\log^5(1-z)$ and $\log^4(1-z)$ for the gluon-gluon channel in eq.~\eqref{eq:ggN3LONSoft} are in agreement 
with the conjecture of ref.~\cite{deFlorian:2014vta}. In ref.\cite{deFlorian:2014vta} a conjecture was also formulated for the colour and flavour structure of the coefficient of $\log^3(1-z)$ up to a rational parameter $\xi^{(3)}_H$. We confirm the validity of this
conjecture for the coefficient of $\log^3(1-z)$ as well and determine $\xi^{(3)}_H=\frac{896}{3}$.  
The $\log^5(1-z)$ coefficient for the quark-gluon channel in eq.~\eqref{eq:qgN3LONSoft} 
agrees with the calculation of ref.~\cite{vogtqg}. The coefficients of the remaining logarithms and the non-logarithmic terms 
in eqs.~\eqref{eq:ggN3LONSoft}-\eqref{eq:qgN3LONSoft} are presented for
the first time in this publication.

\subsection{Coefficients of leading logarithms with exact $z$ dependence}

In this section we obtain another approximation to eq.~\eqref{eq:eta_reg_log}, namely we compute the coefficients of the three leading logarithms in eq.~\eqref{eq:eta_reg_log} with exact $z$ dependence. Indeed, it turns out that the coefficients of these logarithms are uniquely determined at N$^3$LO by requiring the cancellation of the poles in $\eps$, once the single-emission contributions and the counterterms are known.

To be more concrete, we start from eq.~\eqref{eq:etadec} and~\eqref{eq:chi_exp}, and expand all the contributions in the dimensional regulator $\epsilon$,
\begin{eqnarray}
\eta_{ij}^{(3)}(z)&=&\sum\limits_{l=-3}^0\sum\limits_{k=0}^5 \epsilon^{l}\log(1-z)^k \Delta_{ij}^{(3,l,k)}(z)\nonumber\\
&+&\sum\limits_{l=-6}^{0}\epsilon^{l}\left[ \chi_{ij}^{(3,0,l)}\delta(1-z)+\sum\limits_{m=2}^6\chi_{ij}^{(3,m,l)}(z) (1-z)^{-m\epsilon}\right]+\mathcal{O}(\epsilon)\,.
\end{eqnarray}
In order for $\eta_{ij}^{(3)}(z)$ to be finite, all the poles in $\epsilon$ must cancel. This implies that the coefficient of each power of $\log(1-z)$ and of each plus-distribution multiplying a pole in $\epsilon$ has to vanish separately, which allows us to derive a set of equations constraining the individual contributions $\chi^{(3,m,l)}_{ij}(z)$ and $\Delta_{ij}^{(3,l,k)}(z)$. In particular, we get
\beq
\label{eq:logguess}
\Delta^{(3,l,k)}_{ij}(z)+\sum\limits_{m=2}^6 \frac{(-m)^k}{k!} \chi^{(3,m,l-k)}_{ij}(z)=0\,,\hspace{1cm}l<0\,,\,\forall k\,.
\eeq
At this point we note that the terms proportional to $\chi^{(3,2,k)}_{ij}(z)$ and $\chi^{(3,3,k)}_{ij}(z)$ only
receive contributions from single-emission subprocesses, and the computation of those contributions was recently completed for arbitrary values of $z$~\cite{Anastasiou:2013mca,RVV_bf,RVV_ct}. In particular, the computation of the single-emission processes at two loops of ref.~\cite{RVV_bf} has all the logarithms $\log(1-z)$ resummed into factors of the form $(1-z)^{-m\eps}$, which makes the determination of $\chi^{(3,2,k)}_{ij}(z)$ and $\chi^{(3,3,k)}_{ij}(z)$ straightforward.
Including this information we are able to solve the system of equations~\eqref{eq:logguess} for the coefficients of the first three leading logarithms ($\log^{5,4,3}(1-z)$) for all partonic initial states. Parts of the coefficients of these logarithms, corresponding to specific colour coefficients, had already been predicted in ref.~\cite{deFlorian:2014vta}, and we confirm these results. Moreover, we have checked that only the gluon-gluon and quark-gluon initial states give non-vanishing contributions at next-to-soft level, and the values of the coefficients for $z=1$ agree with the corresponding coefficients presented in the previous section.
The analytic results for the different partonic initial states are, for $\mu_R=\mu_F=m_H$,

\begin{eqnarray}
\eta_{gg}^{(3,3),\textrm{reg}}(z)&=&\frac{N_f}{N_c^2}\Bigg[\frac{85}{72} (z+1) H_{1}H_0+\frac{680 z^3-768 z^2-1107 z-276}{864 z} H_{0}\nonumber\\
&&\qquad-\frac{37}{72} (z+1) H_{0}^2-\frac{85}{72} (z+1) H_{2}-\frac{(1-z) \left(2328 z^2+4505 z+1644\right)}{1728 z}\nonumber\\
&&\qquad+\frac{85}{72} (z+1) \zeta_2\Bigg]\nonumber\\ 
 &&+\frac{N_f^{2}}{N_c}\Bigg[-\frac{25}{216} (z+1) H_{0}-\frac{25 (1-z) \left(4 z^2+7 z+4\right)}{1296 z}\Bigg]\nonumber\\ 
 &&+N_f\Bigg[-8 (z+1) H_{1}H_0-\frac{1564 z^3-1229 z^2-2903 z-1820}{432 z} H_{0}\nonumber\\
&&\qquad+\frac{1}{8} (51 z+11) H_{0}^2+8 (z+1) H_{2}+\frac{(1-z) \left(17492 z^2+9035 z+14900\right)}{1296 z}\nonumber\\
&&\qquad-8 (z+1) \zeta_2\Bigg]\nonumber\\ 
 &&+N_c\,N_f^{2}\Bigg[\frac{25}{216} (z+1) H_{0}-\frac{292 z^3-117 z^2+309 z-292}{1296 z}\Bigg]\nonumber\\ 
 &&+N_c^{2}
N_f\Bigg[\frac{491}{72} (z+1) H_{1}H_0-\frac{7184 z^4-19370 z^3+11199 z^2-2377 z+8100}{864 (1-z) z} H_{0}\nonumber\\
&&\qquad+\frac{1}{36} (-211 z-31) H_{0}^2-\frac{491}{72} (z+1) H_{2}\nonumber\\
&&\qquad+\frac{168584 z^3-149895 z^2+172203 z-160268}{5184 z}+\frac{491}{72} (z+1) \zeta_2\Bigg]\nonumber\\ 
 &&+N_c^{3}\Bigg[-\frac{8 \left(z^2+z+1\right)^2 }{z (z+1)}H_{-2}+\frac{8 \left(z^2+z+1\right)^2 }{z (z+1)}H_{-1}H_0-128 (z+1) H_{1}H_0\nonumber\\
&&\qquad+\frac{6259 z^4-13598 z^3+11190 z^2-7514 z+4477}{27 (1-z) z}H_0\nonumber\\
&&\qquad+\frac{2 \left(16 z^5-49 z^4-3 z^3+49 z^2+3 z+14\right) }{(1-z) z (z+1)}H_{0}^2+128 (z+1) H_{2}\nonumber\\
&&\qquad-\frac{19980 z^3-19259 z^2+21100 z-19980}{54 z}\nonumber\\
&&\qquad+\frac{4 \left(15 z^4-30 z^3-47 z^2-16 z-13\right) }{z (z+1)}\zeta_2\Bigg]\,,\\
\eta_{gg}^{(3,4),\textrm{reg}}(z)&=&
\frac{N_f}{N_c^2}\Bigg[\frac{85}{288} (z+1) H_{0}+\frac{85 (1-z) \left(4 z^2+7 z+4\right)}{1728 z}\Bigg]\nonumber\\ 
 &&+N_f\Bigg[-2 (z+1) H_{0}-\frac{(1-z) \left(4 z^2+7 z+4\right)}{3 z}\Bigg]\nonumber\\ 
 &&+N_c^{2}N_f\Bigg[\frac{491}{288} (z+1) H_{0}-\frac{5804 z^3-2367 z^2+6207 z-5804}{1728 z}\Bigg]\nonumber\\ 
 &&+N_c^{3}\Bigg[\frac{671 z^3-641 z^2+751 z-671}{9 z}\nonumber\\
&&\qquad-\frac{27 z^4-86 z^3+81 z^2-22 z+27 }{(1-z) z}H_{0}\Bigg]\,,\\
\eta_{gg}^{(3,5),\textrm{reg}}(z)&=&N_c^{3}\,\frac{8 \left(-z^3+z^2-2 z+1\right)}{z}\,,
 \end{eqnarray}
 
 \begin{eqnarray}
 \label{eq:sqg}
\eta_{qg}^{(3,3),\textrm{reg}}(z)&=&\frac{N_c^2-1}{N_c^3}\Bigg\{\frac{37 z^2-74 z+24 }{256 z}H_{0}^2+\frac{33 z^2-66 z-416 }{1152 z}H_{2}\nonumber\\ 
 &&\qquad+\frac{88 z^3+1813 z^2-2876 z+972}{2304 z}H_0-\frac{\left(33 z^2-66 z-416\right) }{1152 z}H_{1} H_{0}\nonumber\\ 
 &&\qquad-\frac{\left(419 z^2-838 z+356\right) }{1152 z}\zeta_2-\frac{1364 z^3-10401 z^2+21360 z-10424}{13824 z}\nonumber\\ 
 &&+N_c\,N_f\Bigg[-\frac{665 z^2+398 z-1068 }{1728 z}H_{0}+\frac{1}{6} (z-2) H_{0}^2\nonumber\\ 
 &&\qquad-\frac{768 z^3-9403 z^2+16391 z-8414}{5184 z}\Bigg]\nonumber\\ 
 &&+N_c^{2}\Bigg[-\frac{1001 z^2-158 z+732 }{576 z}H_{0}^2+\frac{19 \left(z^2+2 z+2\right) }{144 z}H_{-2}\nonumber\\ 
 &&\qquad-\frac{354 z^2+441 z+37 }{144 z}H_{2}+-\frac{19 \left(z^2+2 z+2\right) }{144 z}H_{-1}H_0\nonumber\\ 
 &&\qquad+\frac{354 z^2+441 z+37 }{144 z}H_{1}H_0+\frac{1896 z^3-25061 z^2+20464 z-26652}{3456 z} H_{0}\nonumber\\ 
 &&\qquad+\frac{1213 z^2-204 z+1084 }{288 z}\zeta_2+\frac{46448 z^3-19855 z^2+318062 z-347740}{20736 z}\Bigg]\nonumber\\ 
 &&+N_c^2N_f^{2}\frac{7 \left(z^2-2 z+2\right)}{864 z}\nonumber\\ 
 &&+N_c^{3}N_f\Bigg[-\frac{593 z^2+230 z+2536 }{1728 z}H_{0}+\frac{1}{6} (2-z) H_{0}^2\nonumber\\ 
 &&\qquad+\frac{1816 z^3-12011 z^2+29119 z-22598}{5184 z}\Bigg]\nonumber\\ 
 &&+N_c^{4}\Bigg[\frac{26023 z^2+4802 z+19608 }{2304 z}H_{0}^2-\frac{125 \left(z^2+2 z+2\right) }{144 z}H_{-2}\nonumber\\ 
 &&\qquad+\frac{5957 z^2+12670 z+3112 }{384 z}H_{2}+\frac{125 \left(z^2+2 z+2\right)}{144 z} H_{-1}H_0\nonumber\\ 
 &&\qquad-\frac{5957 z^2+12670 z+3112 }{384 z}H_{1}H_0\nonumber\\
&&\qquad-\frac{42712 z^3-17807 z^2+131476 z-315012}{6912 z} H_{0}\nonumber\\ 
 &&\qquad-\frac{22953 z^2+25846 z+19500 }{1152 z}\zeta_2\nonumber\\ 
 &&\qquad-\frac{394212 z^3-321247 z^2+3718820 z-3958688}{41472 z}\Bigg] \Bigg\}\,,\\
\eta_{qg}^{(3,4),\textrm{reg}}(z)&=&\frac{N_c^2-1}{N_c^3}\Bigg\{-\frac{565 z^2-1130 z+648 }{4608 z}H_{0}-\frac{2165 z^2-4828 z+2892}{9216 z}\nonumber\\ 
 &&-N_c\,N_f\,\frac{253 \left(z^2-2 z+2\right)}{6912 z}\nonumber\\
 &&+N_c^{2}\Bigg(\frac{742 z^2-335 z+813 }{576 z}H_{0}-\frac{3064 z^3-17033 z^2+58726 z-52316}{13824 z}\Bigg)\nonumber\\
&&+N_c^{3}\,N_f\frac{803 \left(z^2-2 z+2\right)}{6912 z}\nonumber\\
&&+N_c^{4}\Bigg(\frac{49168 z^3+8689 z^2+388000 z-455012}{27648 z}\nonumber\\
&&\qquad-\frac{9929 z^2+4726 z+11056 }{1536 z}H_{0}\Bigg)\Bigg\}\,,\\
\eta_{qg}^{(3,5),\textrm{reg}}(z)&=&\frac{N_c^2-1}{N_c^3}\,\frac{z^2-2 z+2}{z}\,\Bigg(\frac{9}{256}-N_c^{2}\,\frac{77}{384}+N_c^{4}\,\frac{587}{768}\Bigg)\,,\\
%
  \label{eq:sqqb}
\eta_{q\bar{q}}^{(3,3),\textrm{reg}}(z)&=&\frac{(N_c^2-1)^2}{N_c^4}\Bigg\{\frac{10 z^3-39 z^2+39 z-16 }{48 z}H_{0}-\frac{(1-z) \left(121 z^2-206 z+121\right)}{288 z}\nonumber\\ 
 &&+N_c\Bigg[-\frac{73 z^2+292 z+196 }{384 z}H_{0}^2+ \frac{13 (z+2)^2 }{24 z}H_{1}H_{0}+\frac{71 z^2-160 z-1092}{384 z}H_{0}\nonumber\\ 
 &&\qquad-\frac{13 (z+2)^2 }{24 z}H_{2}+\frac{13 (z+2)^2 }{24 z}\zeta_2-\frac{(1-z) (569 z+1301)}{256 z}\Bigg]\nonumber\\ 
 &&+N_c\,N_f\frac{11 (1-z)^3}{216 z}\nonumber\\ 
 &&+N_c^{2}\Bigg[\frac{(1-z) \left(247 z^2-521 z+247\right)}{216 z}-\frac{34 z^3-93 z^2+93 z-28 }{48 z}H_{0}\Bigg]\nonumber\\ 
 &&+N_c^2N_f\Bigg[-\frac{(z+2)^2 }{48 z}H_{0}-\frac{(1-z) (z+3)}{24 z}\Bigg]\nonumber\\ 
 &&+N_c^{3}\Bigg[\frac{347 z^2+236 z+908 }{384 z}H_{0}^2-\frac{35 (z+2)^2 }{24 z}H_{1}H_{0}-\frac{1193 z^2-496 z-4308}{384 z}H_{0}\nonumber\\ 
 &&\qquad+\frac{35 (z+2)^2 }{24 z}H_{2}+\frac{(1-z) \left(512 z^2+95 z+48419\right)}{2304 z}-\frac{35 (z+2)^2 }{24 z}\zeta_2\Bigg]\nonumber\\
 &&+N_c^3N_f\,\frac{19 (1-z)^3}{216 z}\nonumber\\ 
 &&+N_c^{4}\Bigg[\frac{5 z^3-12 z^2+12 z-3 }{12 z}H_{0}-\frac{(1-z) \left(1309 z^2-2762 z+1309\right)}{864 z}\Bigg]\Bigg\}\,,\\
\eta_{q\bar q}^{(3,4),\textrm{reg}}(z)&=&\frac{(N_c^2-1)^2}{N_c^4}\Bigg\{-\frac{7 (1-z)^3}{192 z}+N_c\Bigg[\frac{13 (z+2)^2 }{96 z}H_{0}+\frac{13 (1-z) (z+3)}{48 z}\Bigg]\nonumber\\
&&-N_c^{2}\frac{7 (1-z)^3}{48 z}+N_c^{3}\Bigg[-\frac{35 (z+2)^2 }{96 z}H_{0}-\frac{35 (1-z) (z+3)}{48 z}\Bigg]\nonumber\\ 
 &&+N_c^{4}\,\frac{35 (1-z)^3}{192 z}\Bigg\}\,,\\
%
 \label{eq:sqq}
\eta_{q q}^{(3,3),\textrm{reg}}(z)&=&\frac{(N_c^2-1)^2}{N_c^4}\Bigg\{N_c\Bigg[-\frac{73 z^2+292 z+196 }{384 z}H_{0}^2+ \frac{13 (z+2)^2 }{24 z}H_{1}H_{0}+\frac{13 (z+2)^2 }{24 z}\zeta_2\nonumber\\
&&\qquad+\frac{71 z^2-160 z-1092}{384 z}H_{0}-\frac{13 (z+2)^2 }{24 z}H_{2}-\frac{(1-z) (569 z+1301)}{256 z}\Bigg]\nonumber\\ 
 &&+N_c^{2}N_f\Bigg[-\frac{(z+2)^2 }{48 z}H_{0}-\frac{(1-z) (z+3)}{24 z}\Bigg]\nonumber\\ 
 &&+N_c^{3}\Bigg[\frac{347 z^2+236 z+908 }{384 z}H_{0}^2-\frac{35 (z+2)^2 }{24 z}H_{1}H_{0}-\frac{1193 z^2-496 z-4308}{384 z}H_{0}\nonumber\\
 &&\qquad+\frac{35 (z+2)^2 }{24 z}H_{2}+\frac{(1-z) \left(512 z^2+95 z+48419\right)}{2304 z}-\frac{35 (z+2)^2 }{24 z}\zeta_2\Bigg]\Bigg\}\,,\\
\eta_{q q}^{(3,4),\textrm{reg}}(z)&=&\frac{(N_c^2-1)^2}{N_c^4}\Bigg\{N_c\Bigg[\frac{13 (z+2)^2 }{96 z}H_{0}+\frac{13 (1-z) (z+3)}{48 z}\Bigg]\nonumber\\ 
 &&+N_c^{3}\Bigg[-\frac{35 (z+2)^2 }{96 z}H_{0}-\frac{35 (1-z) (z+3)}{48 z}\Bigg]\Bigg\}\,,\\
%
 \label{eq:sqqp}
\eta_{q q^{\prime}}^{(3,3),\textrm{reg}}(z)&=&\frac{(N_c^2-1)^2}{N_c^4}\Bigg\{N_c\Bigg[-\frac{73 z^2+292 z+196 }{384 z}H_{0}^2+\frac{13 (z+2)^2 }{24 z}H_{1}H_{0}\nonumber\\
&&\qquad +\frac{71 z^2-160 z-1092}{384 z}H_{0} -\frac{13 (z+2)^2 }{24 z}H_{2}+\frac{13 (z+2)^2 }{24 z}\zeta_2\nonumber\\
&&\qquad-\frac{(1-z) (569 z+1301)}{256 z}\Bigg]\nonumber\\ 
 &&+N_c^{2}N_f\,\Bigg[-\frac{(z+2)^2 }{48 z}H_{0}-\frac{(1-z) (z+3)}{24 z}\Bigg]\nonumber\\ 
 &&+N_c^{3}\Bigg[\frac{347 z^2+236 z+908 }{384 z}H_{0}^2-\frac{35 (z+2)^2 }{24 z}H_{1}H_{0}\nonumber\\
 &&\qquad -\frac{1193 z^2-496 z-4308}{384 z}H_{0} +\frac{35 (z+2)^2 }{24 z}H_{2}\nonumber\\
 &&\qquad+\frac{(1-z) \left(512 z^2+95 z+48419\right)}{2304 z}-\frac{35 (z+2)^2 }{24 z}\zeta_2\Bigg]\Bigg\}\,,\\
\eta_{q q^{\prime}}^{(3,4),\textrm{reg}}(z)&=&\frac{(N_c^2-1)^2}{N_c^4}\Bigg\{N_c\Bigg[\frac{13 (z+2)^2 }{96 z}H_{0}+\frac{13 (1-z) (z+3)}{48 z}\Bigg]\nonumber\\ 
 &&+N_c^{3}\Bigg[-\frac{35 (z+2)^2 }{96 z}H_{0}-\frac{35 (1-z) (z+3)}{48 z}\Bigg]\Bigg\}.
  \end{eqnarray}
Note that $\eta_{q \bar q}^{(3,5),\textrm{reg}}(z) = \eta_{q q}^{(3,5),\textrm{reg}}(z) = \eta_{q q^{\prime}}^{(3,5),\textrm{reg}}(z) =0$. We have written the results in terms of harmonic polylogarithms~\cite{Remiddi:1999ew}
 \beq\bsp
 H_{0}&\,=\log z\,, \\
 H_{1}&\,=-\log(1-z)\,,\\
 H_{-1}&\,=\log(1+z)\,,\\
 H_{2}&\,=\text{Li}_2(z)\,,\\
 H_{-2}&\,=-\text{Li}_2(-z)\,.
\esp\eeq

\section{Numerical results for the N$^3$LO hadronic cross-section}
\label{sec:numerics}
In this Section, we  will study the numerical impact of the
partonic N$^3$LO corrections of Section~\ref{sec:results} on the hadronic
Higgs-boson production cross-section.  We normalise all our
results to the leading-order hadronic cross-section, and we factor out the
Wilson coefficient (i.e., we set $C=1$). We choose the Higgs-boson mass to be
$m_H=125 {\rm GeV}$ and compute the cross-sections for a proton-proton
collider with a center-of-mass energy of $14 {\rm TeV}$.  We use the MSTW2008 NNLO parton
densities for all orders and the corresponding value of $\alpha_s(M_Z)$~\cite{mstw}. We set the
renormalisation and factorisation scales equal to the Higgs-boson mass, $\mu_R=\mu_F = m_H$.

\subsection{Results in the threshold expansion}
We start our numerical analysis by studying the behavior of the hadronic cross-section at N$^3$LO through the first two terms in the threshold expansion. 
For assessing the numerical importance of the corrections, it is useful to  substitute the number of colours and number of light quark flavours by their physical values ($N_c=3,N_f=5$ respectively) into 
eq.~\eqref{eq:ggN3LONSoft} and~\eqref{eq:qgN3LONSoft}.  We find,
\begin{align}
\label{eq:ggN3LONSoft_num}
 \eta_{gg}^{(3)}(z)_{\big|(1-z)^0}  &= 
-256 \log^5(1-z) \quad & (\to \quad 115.33\%)
\nonumber \\ &
+959 \log^4(1-z) \quad & (\to \quad 101.07\%)
\nonumber \\ &
+1254.029198\ldots \log^3(1-z) \quad & (\to \quad -32.15\%)
\nonumber \\ &
- 11089.328274\ldots   \log^2(1-z) \quad & (\to \quad -89.41\%)
\nonumber \\ &
+ 15738.441212\ldots  \log(1-z) \quad & (\to \quad -55.50\%)
\nonumber \\ &
-5872.588877\ldots \quad & (\to \quad -14.31\%)
\end{align}
and 
\begin{align}
\label{eq:qgN3LONSoft_num}
\eta_{qg}^{(3)}(z)_{\big|(1-z)^0}  =& 
\frac{1283}{72} \log^5(1-z)  \quad & (\to \quad -14.74\%)
\nonumber \\ &
-\frac{5215}{2592} \log^4(1-z) \quad & (\to \quad -0.33\%)
\nonumber \\ &
- 114.569021\ldots  \log^3(1-z) \quad & (\to \quad 4.58\%)
\nonumber \\ &
 + 513.562980\ldots   \log^2(1-z) \quad & (\to \quad 6.51\%)
\nonumber \\ &  
- 313.985230\ldots \log(1-z) \quad & (\to \quad 1.77\%)
\nonumber \\ &
+ 204.620790\ldots \quad & (\to \quad 0.83\%).
\end{align}
In parentheses we show the relative size of the correction which each term induces to the hadronic 
cross-section relatively to  the leading order contribution from
$\eta_{gg}^{(0)} = \delta(1-z)$.

We find that the formally most singular terms cancel against less
singular ones.  In addition to the large cancellations among different
powers of logarithms, we notice that the formal hierarchy of their
magnitude does not correspond to a similar hierarchy at the hadronic
cross-section level. These observations are the same as we had
already noted in ref.~\cite{Anastasiou:2014vaa} for the leading terms of the soft
expansion.  For ease of comparison, we also recite here the 
analogous decomposition of the leading terms in the soft expansion~\cite{Anastasiou:2014vaa}
\begin{align}
\label{eq:ggN3LOSoft_num}
\eta_{gg}^{(3)}(z)&\simeq 
\Dplus{5}\,216\,.                                & (\to  93.72\%) \nonumber \\ 
 &-\Dplus{4}\,230                                 & (\to  20.01\%) \nonumber \\
& -\Dplus{3}\,1824.362531\ldots        & (\to  -39.90\%)\nonumber\\
& + \Dplus{2}\,7116.015302\ldots      & (\to  -52.45\%)\nonumber\\
& - \DplusOne\,6062.086738\ldots    & (\to  -22.88\%)\nonumber\\
& + \DplusZero\,1466.478272\ldots   & (\to  -5.85\%) \nonumber\\
& +\delta(1-z)\,1124.308887\ldots          & (\to   5.1\%).
\end{align}
The total contribution of
$\eta_{gg}^{(3)}(z)_{\big|(1-z)^0}$ to the hadronic
cross-section is about $25\%$ of the Born contribution, while the
contribution of $\eta_{qg}^{(3)}(z)_{\big|(1-z)^0}$ is about
$-1.38\%$ of the Born contribution. This has to be contrasted with the leading soft 
contribution at N$^3$LO from $\eta_{gg}^{(3)}(z)_{\big|(1-z)^{-1}}
$ which is only $-2.25\%$ of the Born. While the next-to-soft
correction for kinematics corresponding to threshold production should
be suppressed, instead it turns out to be much larger than the leading
threshold contribution.

It is often preferred in the literature to perform the threshold expansion in
Mellin space. The Mellin transformation of a function $f(z)$ is defined as
\begin{equation}
M[f](N) = \int_0^1 dz\, z^{N-1}\, f(z)\,.
\end{equation}
The Mellin transformation is invertible, and the inverse transformation reads 
\begin{equation}
M^{-1}\left[g\right](z) = \int_{c-i\infty}^{c+i \infty} \frac{dN}{2\pi i}\, g(N)\, x^{-N}\,, 
\end{equation}
where the real part of $c$ is chosen such that the poles of $g(N)$ lie to the left of the 
integration contour. One of the main properties of the Mellin transformation is that it maps convolutions as in eq.~\eqref{eq:convolution} to the product of the Mellin transformations,
\beq
M[A\otimes B](N) = M[A](N)\,M[B](N)\,.
\eeq
It follows that the convolution of the partonic cross-sections with the parton densities factorises and turns into an ordinary product in Mellin space. 
Hence, in order to compute the Mellin transformation of the
total hadronic cross-section, we need the Mellin transformations of the parton densities. To this effect, we fit the parton densities for a fixed scale to a functional form
of the type 
\[ 
f_i(x) = x^{a_i} (1-x)^{b_i} (c_{i,0}+c_{i,1} x+c_{i,2} x^2 +\ldots)\,,
\]
for which we can easily compute the Mellin transformation using Euler's Beta function,
\begin{equation}
\label{eq:MTbetafunction}
M\left[x^a (1-x)^b\right](N) = \frac{\Gamma(N+a)\Gamma(1+b)}{\Gamma(1+a+b+N)}.
\end{equation}
For the partonic cross-section we perform an expansion around the
threshold limit, which in Mellin space
corresponds to taking $N \to \infty$.  Through $\ord(\frac{1}{N})$, we
find: 
\begin{align}
\label{eq:ggN3LOsvns_mellin}
M\left[\eta_{gg}^{(3)}\right](N)\simeq 
 \,&36\log^6N	& (\to  0.0013\%) \nonumber\\ 
&+170.679\dots\log^5N	& (\to  0.0226\%) \nonumber\\ 
&+744.849\dots\log^4N	& (\to  0.2570\%) \nonumber\\ 
&+1405.185\dots\log^3N	& (\to  1.0707\%) \nonumber\\ 
&+2676.129\dots\log^2N	& (\to  4.0200\%) \nonumber\\ 
&+1897.141\dots\log N	& (\to  5.1293\%) \nonumber\\ 
&+1783.692\dots			& (\to  8.0336\%) \nonumber\\ 
&+108\frac{\log^5N}{N}				& (\to  0.0105\%) \nonumber\\
&+615.696\dots\frac{\log^4N}{N}	& (\to  0.1418\%) \nonumber\\
&+2036.407\dots\frac{\log^3N}{N}	& (\to  0.9718\%) \nonumber\\
&+3305.246\dots\frac{\log^2N}{N}	& (\to  2.9487\%) \nonumber\\
&+3459.105\dots\frac{\log N}{N}		& (\to  5.2933\%) \nonumber\\
&+703.037\dots\frac{1}{N}			& (\to  1.7137\%).
\end{align}
In parentheses we show the relative size of the correction which each term induces to the hadronic 
cross-section relatively to  the leading order contribution from
$\eta_{gg}^{(0)} = \delta(1-z)$. 
In Mellin space the pattern of corrections in the threshold expansion is different from the one observed in $z$-space.
As it was also observed for the leading soft terms and parts of the next-to-soft terms 
in ref.~\cite{deFlorian:2014vta}, we find that through $\ord{\left(\frac 1 N  \right) }$ the corrections are always positive. 
Nevertheless, we observe that the formally leading logarithms contribute the 
least to the hadronic cross-section.  In total, the soft-virtual (SV) terms
($\log^nN$) contribute about $\sim 18\%$ of the Born to the cross-section, 
while the next-to-soft (NS) terms  (${\log^nN}/{N}$) contribute
about $\sim 11\%$ of the Born.  
We therefore conclude that, unlike common folklore suggests, the 
threshold limit does in fact not dominate the cross-section at LHC energies, but there is a sizeable contribution from terms beyond threshold.

As we have emphasised in ref.~\cite{Anastasiou:2014vaa}, there is an ambiguity in 
how to convolute an approximate partonic cross-section with the parton densities. 
For example, we can recast the  hadronic cross-section in the form,
\begin{equation}
\label{eq:sigma_alpha}
\sigma = \tau^{1+n} \sum_{ij} \left( f_i^{(n)} \otimes f_j^{(n)}  \otimes 
\frac{\hat{\sigma}_{ij}(z)}{z^{1+n}}
\right)\left(\tau \right) 
\end{equation}
where 
\begin{equation}
f_i^{(n)}(z) \equiv \frac{f_i(z)}{z^n}.
\end{equation}
$\sigma$ is independent of the arbitrary parameter $n$ as long as the partonic cross-section 
is known exactly. 
Mellin transforming eq.~\eqref{eq:sigma_alpha}, we obtain
\beq\bsp
M\left[\frac{\sigma}{\tau^{1+n}}\right](N) &\,= \sum_{ij} M\left[f^{(n)}_i\right](N) \, M\left[f^{(n)}_j\right](N)\, M\left[\frac{\hat{\sigma}(z)}{z^{1+n}}\right](N)\\
&\,=\sum_{ij} M\left[f_i\right](N-n) \, M\left[f_j\right](N-n)\, M\left[\frac{\hat{\sigma}(z)}{z}\right](N-n)\,.
\esp\eeq
If only a finite number of terms in the threshold expansion of the partonic cross-sections
are kept,
\begin{equation}
\label{eq:n-variation}
\frac{\hat{\sigma}_{ij}(z)}{z^{1+n}} \simeq  
\left. \hat{\sigma}_{ij}(z) \right|_{(1-z)^{-1}} + \left. \hat{\sigma}_{ij}(z) \right|_{(1-z)^{0}}+n (1-z) \left. \hat{\sigma}_{ij}(z) \right|_{(1-z)^{-1}} + 
\ord{(1-z)^1}
\end{equation}
then the convolution integral is sensitive to varying the
arbitrary parameter $n$. 
This ambiguity is expected to be reduced when including higher-order terms in the threshold expansion. 
This effect was already observed at NNLO~\cite{nnlosoft},
corresponding to expanding around threshold the $1/z$ flux-factor as
part of the partonic cross-section or evaluating it unexpanded as
part of the parton luminosity. A similar ambiguity appears to be
responsible~\cite{Sterman:2013nya,Bonvini:2014qga} for the bulk of the difference in the numerical
predictions for the Higgs cross-section at N$^3$LO in various approaches and implementations of 
threshold resummation~\cite{nnloresummation}.


In the remainder of this section we analyse the impact of this truncation when we use the results of Section~\ref{sec:results}, which contains the most precise information on the threshold expansion of the cross-section at N$^3$LO to date. In order to quantify the trustworthiness of the threshold approximation, we study the dependence of the result on the parameter $n$ defined through eq.~\eqref{eq:sigma_alpha}, both in $z$ and in Mellin-space.

 \begin{figure}[!t]
\includegraphics[width=1.0\textwidth]{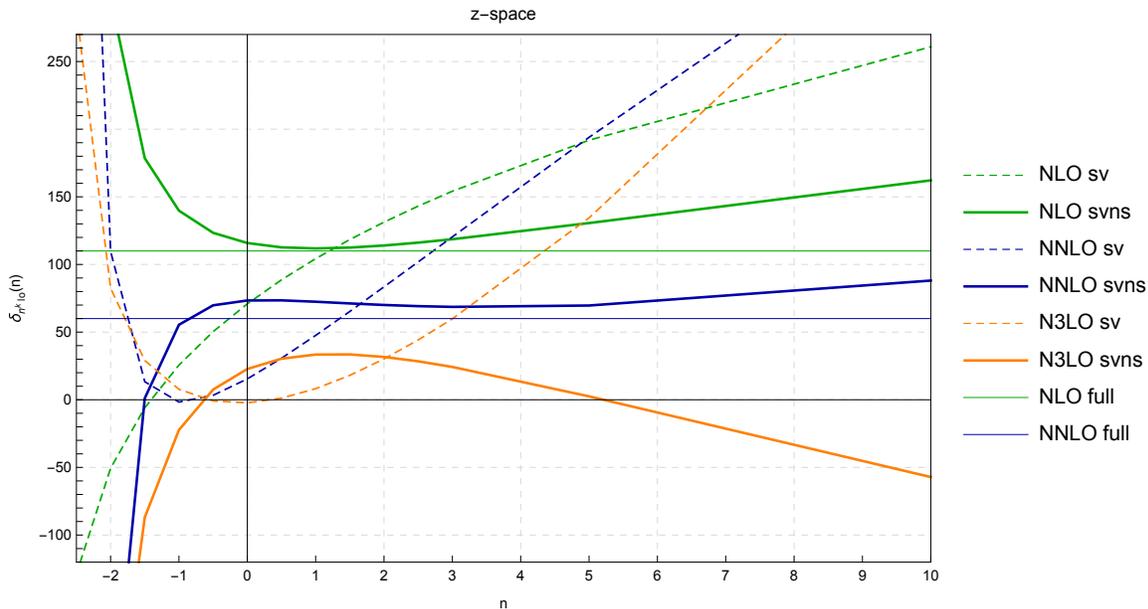}
\caption{
Soft-virtual and next-to-soft corrections at NLO, NNLO and N$^3$LO 
normalised to the Born cross-section in $z-$space as a function of the
artificial parameter $n$ in eq.~\eqref{eq:n-variation}}
\label{fig:Zspacevariation}
\end{figure}

\begin{figure}[!t]
\includegraphics[width=1.0\textwidth]{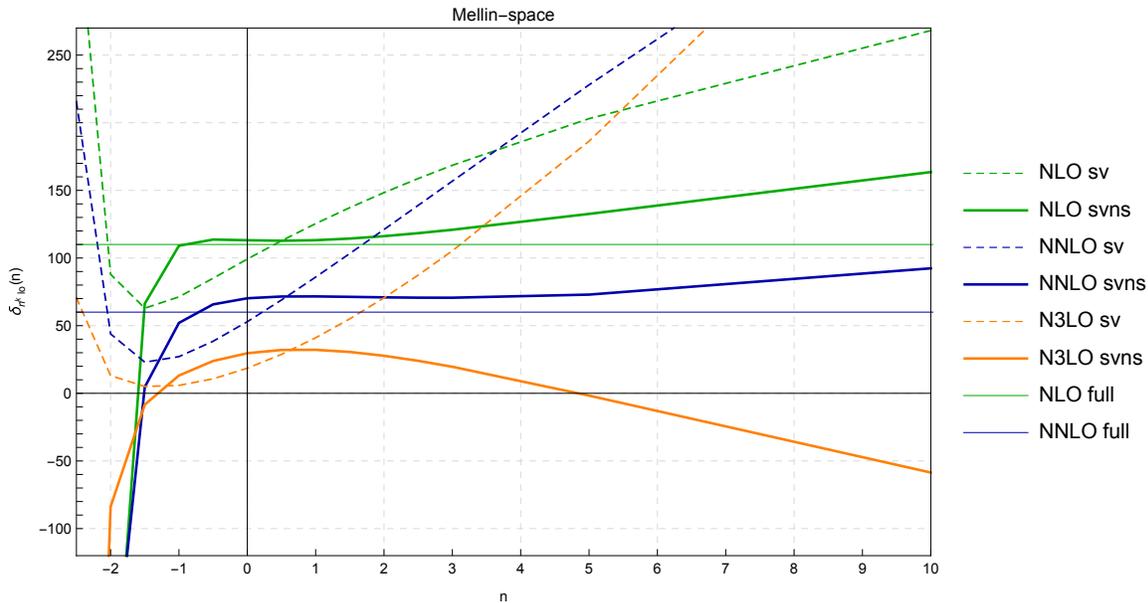}
\caption{Soft-virtual and next-to-soft corrections at NLO, NNLO and N$^3$LO 
normalised to the Born cross-section in Mellin space as a function of the
artificial parameter $n$ in eq.~\eqref{eq:n-variation}}
\label{fig:Nspacevariation}
\end{figure}

In Fig.~\ref{fig:Zspacevariation} we plot the soft-virtual and next-to-soft corrections at NLO, NNLO and N$^3$LO 
normalised to the Born cross-section in $z-$space as a function of the
artificial parameter $n$ in eq.~\eqref{eq:n-variation}.  
In Fig.~\ref{fig:Nspacevariation} we plot the soft-virtual and next-to-soft corrections at NLO, NNLO and N$^3$LO 
normalised to the Born cross-section in Mellin space as a function of the
artificial parameter $n$ in eq.~\eqref{eq:n-variation}. 
We also plot in both figures the known NLO and NNLO corrections as straight lines since they are
insensitive to the  value of $n$.  The full NLO corrections are about
$110\%$ of the Born and the full NNLO corrections are about $60\%$. 
The  sensitivity of the `leading soft' corrections to $n$ is large
at all perturbative orders and in both spaces. This
sensitivity is reduced when the next-to-soft terms are included,
where a plateau at NLO and NNLO is formed for values of $n$ larger than 
about $-1$ and up to very large positive values of $n$. While an improved convergence is
visible, at N$^3$LO the
sensitivity of the next-to-soft correction in $n$ is enhanced in comparison
to NLO and NNLO and there is much less of a plateau.  
The increased sensitivity of the truncated expansion to the artificial parameter $n$ is a symptom 
of the fact that the threshold limit is less dominant at higher orders.  In Table~\ref{tab:NSoverSV} 
we present the ratio of the NS over the SV contribution in the gluon-gluon channel 
(this ratio is infinite in all other channels) both in Mellin and $z-$space.  We observe that the ratio increases 
at higher perturbative orders and hence the soft approximation is increasingly untrustworthy.  This behavior is particularly pronounced in 
$z-$space. 

\begin{table}[h]
\begin{center}
$
\begin{array}{c|rrrrrr}
\hline\hline
 &\phantom{aa}&  {\rm NLO}  &\phantom{aa}& {\rm NNLO} &\phantom{aa}& {\rm N}^3{\rm LO}     \\
\hline
z-{\rm space} & &63.42 \% && 376.5 \%  && -1106.5\%  \\ 
{\rm Mellin-space} & & 14.02 \% && 32.71\% && 59.78 \%  \\ 
\hline\hline
\end{array}
$
\end{center}
\caption{\label{tab:NSoverSV}The ratio of the next-to-soft and the soft-virtual  contribution in Mellin and $z-$space for 
$n=0$ at NLO, NNLO and N$^3$LO.}
\end{table}

Is it possible to use the soft-virtual~\cite{Anastasiou:2014vaa} or the next-to-soft
approximation presented in this article in order to estimate precisely
the N$^3$LO corrections to the Higgs cross-section?  The  fact that the
soft expansion does not yet appear to be convergent, as we discussed
above, does not justify such attempts theoretically.  Nevertheless,
efforts have been made in the literature to guess the full N$^3$LO
corrections from available or estimated soft terms using empirical arguments based on 
the experience from the behavior of the NLO and NNLO corrections.  
The level of precision which must be achieved with empirical
estimations should be better than the $\sim \pm 4\%$ N$^3$LO 
scale variation ~\cite{Buehler:2013fha}  which corresponds to 
$\pm 12\%$ of the Born (the normalization of our
plots). We do not believe that empirical arguments should replace 
proper convergence criteria. However, if we entertain the idea
that a guess can be made by comparing the soft terms with the full
result at NLO and NNLO, we see that the next-to-soft approximation
for $n \in \left[-1,  3\right]$ is close to the full result at NLO ($110\%$ of the Born) and
NNLO ($60\%$ of the Born) in both $z-$space and Mellin space, with an envelope of predictions ranging from $109\%$ to $140\%$ of the Born at NLO 
and from $52\%$ to $73\%$ of the Born at NNLO. 
At N$^3$LO, the variation of the cross-section in both spaces for the same range of $n$ is   from
$-22\%$ to $33\%$ of the Born, which is larger than the target precision at that order.

\subsection{Results for the $\log^{5,4,3}(1-z)$ terms in full kinematics}
 
\begin{table}[h]
\begin{center}
$
\begin{array}{c|rrrrrrrrr}
\hline\hline
 & gg &\phantom{aa}& qg &\phantom{aa}& q \bar q &\phantom{aa}& qq &\phantom{aa}& qQ   \\
\hline
\log^5(1-z) & 111.908\% && -15.006\% && 0\% && 0\% &&  0\% \\ 
\log^4(1-z) &  93.868\% && -0.780\% && 0.002\% && 0.002\% && 0.009\%  \\ 
\log^3(1-z) & -39.201\% && 3.459\% && 0.004\% && 0.003\% && 0.017\%  \\  \hline\hline
\end{array}
$
\end{center}
\caption{\label{tab:fulllogs}The  contribution of the $\log^{5,4,3}(1-z)$ in full kinematics to the hadronic cross-section normalized to the Born, for each 
partonic channel.}
\end{table}

It is clear from the above that a reliable estimate of the N$^3$LO correction of the 
Higgs cross-section requires even more terms in the threshold expansion. As explained 
in Sections~\ref{sec:results}, we have been able to obtain the coefficients of the $\log^{5,4,3}(1-z)$
terms in a closed form, valid for arbitrary values of $z$.  These corrections are insensitive to 
the artificial parameter $n$, and thus independent of whether we perform the computation in Mellin or $z-$space.  
Their contribution to the hadronic cross-section from each partonic channel normalized to the Born hadronic 
cross-section (setting the Wilson coefficient $C=1$) is shown in Table~\ref{tab:fulllogs}. 
Comparing the effect of the full $\log^{5,4,3}(1-z)$ coefficients to the truncated ones  in the $(1-z)$ expansion, 
as in eq.~\eqref{eq:ggN3LONSoft_num} and~\eqref{eq:qgN3LONSoft_num}, we find that the full coefficients give systematically lower
contributions to the hadronic cross-section. 

\begin{figure}[!t]
\includegraphics[width=\textwidth]{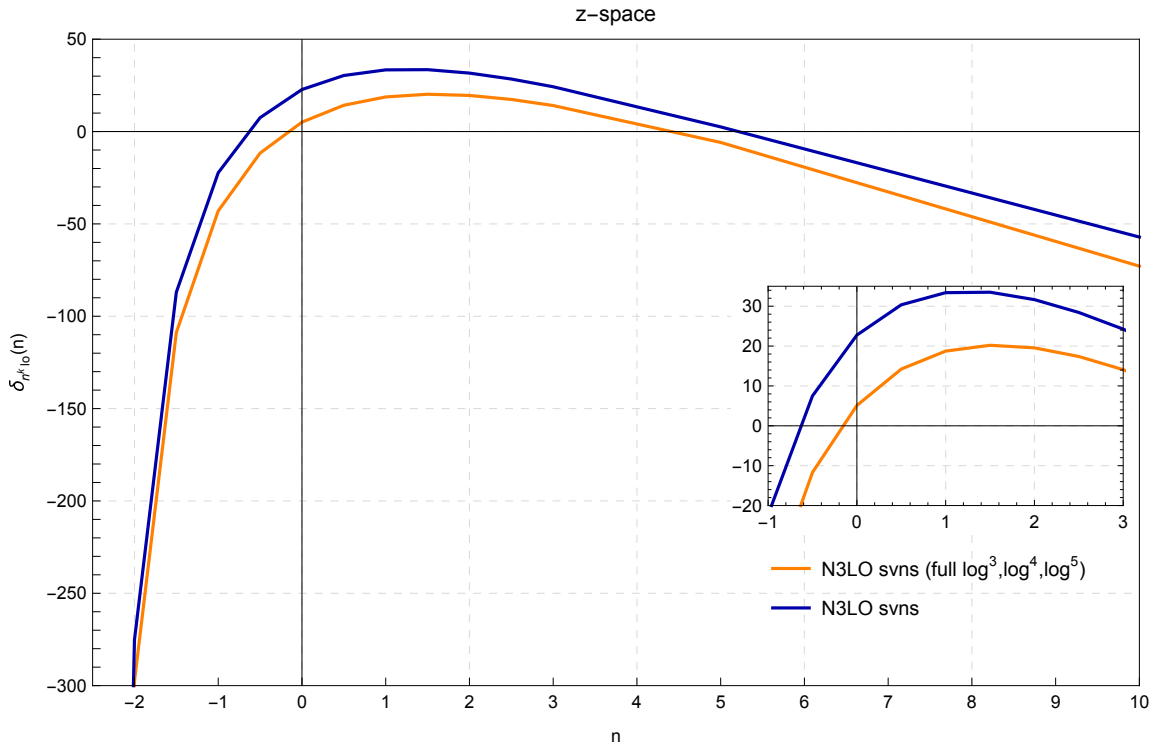}
\caption{ The hadronic cross-section at N$^3$LO where  the $\log^{5,4,3}(1-z)$ and $\delta(1-z)$ contributions are 
computed in full kinematics while the  remaining $\log^{2,1,0}(1-z)$ terms are computed in the soft-virtual and next-to-soft 
approximation in $z-$space as a function of the 
artificial parameter $n$. 
The cross-section is normalized to the Born cross-section and only the 
dominant $gg$-channel is included. 
}
\label{fig:Zfspacevariation}
\end{figure}
Knowing the exact $\log^{5,4,3}(1-z)$ coefficients, we can restrict the threshold approximation only to the 
coefficients of the $\log^{2,1,0}(1-z)$ terms.  This mixed approach would not have been justified if we had found that the formal threshold expansion hierarchy 
was reflected in the results after the integration over the parton densities. However, this is not the case and it is therefore equally justified (or unjustified) to include the full 
kinematic dependence of the coefficients of the `leading' logarithms.  We present in Fig.~\ref{fig:Zfspacevariation} the corresponding gluon-channel contribution to the 
hadronic cross-section normalised to the Born cross-section, as a function of the artificial exponent $n$. As expected from the comparison of the results of Table~\ref{tab:fulllogs} in full 
kinematics and the results of eqs.~\eqref{eq:ggN3LONSoft} in the threshold expansion for the $\log^{5,4,3}(1-z)$ terms,  the inclusion of the full leading logarithms 
lowers the value of the N$^3$LO correction. The shape as a function of $n$, however, does not substantially change. This indicates that the bulk of the $n$ dependence is carried by the coefficients of the yet-unknown $\log^{2,1,0}(1-z)$ terms, and including the exact coefficients of $\log^{5,4,3}(1-z)$ does not substantially improve the convergence of the threshold expansion. It is unclear whether the inclusion of the yet unknown full coefficients for the $\log^{2,1,0}(1-z)$ terms in the future will further reduce or increase the cross-section. 
 \begin{figure}[!t]
\includegraphics[width=1.0\textwidth]{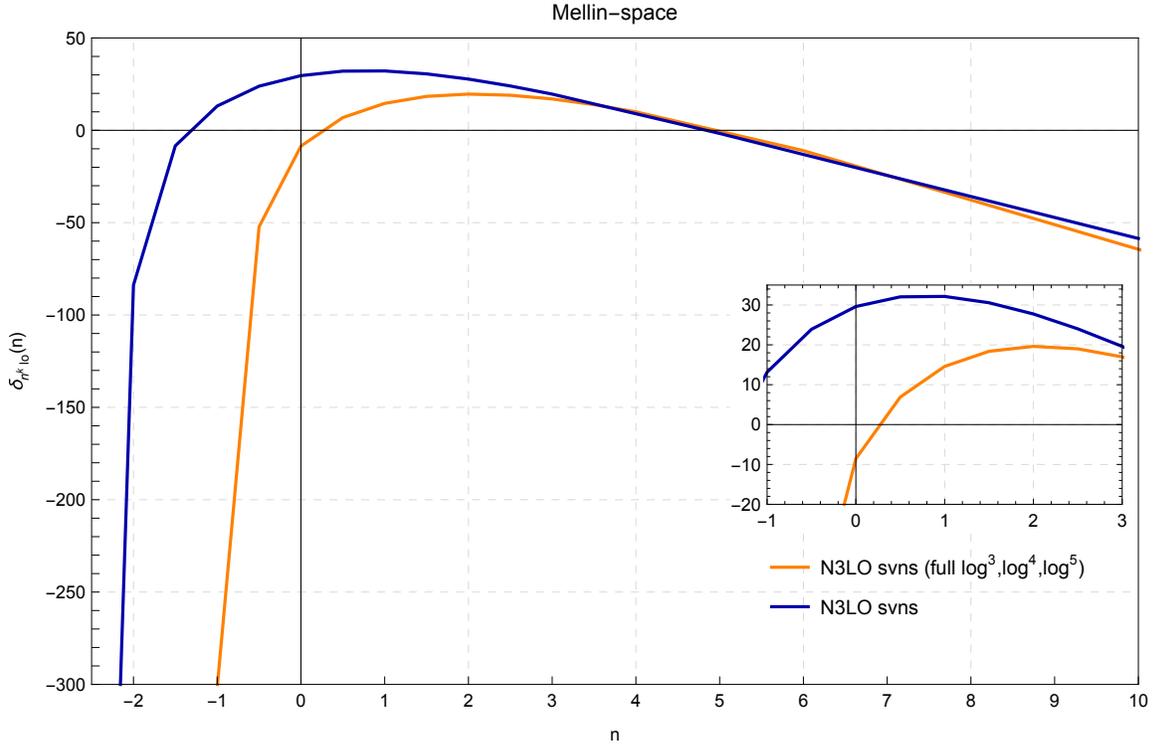}
\caption{ The hadronic cross-section at N$^3$LO where  the $\log^{5,4,3}(1-z)$ and $\delta(1-z)$ contributions are 
computed in full kinematics while the  remaining $\log^{2,1,0}(1-z)$ terms are computed in the soft-virtual and next-to-soft
approximation in Mellin space as a function of the 
artificial parameter $n$ in Eq.~\ref{eq:n-variation}. 
The cross-section is normalized to the Born cross-section and only the 
dominant $gg$-channel is included. 
}
\label{fig:Nfspacevariation}
\end{figure}
In Fig.~\ref{fig:Nfspacevariation}, we include the full $\log^{5,4,3}(1-z)$ terms exactly and 
compute the remaining known N$^3$LO terms as a threshold expansion in Mellin space. The reduction of the cross-section 
is even more pronounced in this case. For example, setting $n=0$, the pure next-to-soft approximation 
in Mellin space yields a positive contribution of about $+29.5\%$ of the Born, while including the exact contribution from $\log^{5,4,3}(1-z)$ 
and expanding in Mellin space the remaining terms through next-to-soft yields a negative N$^3$LO correction of about $-8.5\%$ of the Born.  

The changes that we observe by including the full coefficients of $\log^{5,4,3}(1-z)$ with respect to pure next-to-soft approximations have to be compared with smaller 
scale variation uncertainty at N$^3$LO ~\cite{Buehler:2013fha}, which is about $\pm 12\%$ of the Born cross-section. While in this publication we have presented 
the most advanced theoretical calculation of the N$^3$LO corrections, we conclude that this is insufficient to reduce the theoretical uncertainty of the Higgs-boson cross-section.

\section{Conclusions}
\label{sec:conclusions}
In this paper we have presented new results for Higgs-boson production at N$^3$LO beyond threshold. More precisely, we have computed for the first time the full next-to-soft corrections to Higgs-boson production, as well as the exact results for the coefficients of the first three leading logarithms at N$^3$LO. Our results constitute a major milestone towards the complete computation of the Higgs-boson cross-section via gluon-fusion at N$^3$LO.

Having at our disposal the formally most accurate result for the threshold expansion available to date, we are naturally lead to the question of how reliable phenomenological predictions based on this result would be. In a second part of our paper we therefore performed a critical appraisal of the threshold approximation, which according to the general folklore captures the bulk of the Higgs-boson cross-section. Unfortunately, the convergence of the threshold expansion appears to become less reliable with each further order in the perturbative expansion, as formally subleading terms are not suppressed in comparison to leading 
terms. In this context, we make the alarming observation that the ratio of the next-to-soft over the soft-virtual corrections increases from NLO to NNLO and to N$^3$LO showing that the threshold approximation deteriorates when applied to higher orders in the perturbative QCD expansion.

A second problem in using the threshold expansion is that there is an ambiguity in defining the convolution integral for the hadronic cross-section 
from the threshold expansion of the partonic cross-sections. We have introduced in eq.~\eqref{eq:sigma_alpha} a way to quantify this ambiguity by introducing a parameter $n$ such that the hadronic cross section is independent of $n$ if no approximation is made. The truncation of the threshold expansion, however, introduces a dependence on $n$, and the size of this dependence is a measure for the convergence of the threshold expansion. We have performed a numerical study of the $n$-dependence by including terms beyond the strict threshold limit, both in $z$-space and in Mellin-space. We observe that in all cases the numerical dependence on $n$ is decreased when including corrections beyond threshold, in agreement with the expectations. 
At NLO and NNLO, a plateau (numerically close to the true value) forms when next-to-soft terms are included. 
At N$^3$LO, however, we observe that no plateau is visible, indicating that empirical estimations of the 
N$^3$LO cross-section based on the experience from NLO and NNLO may fail. In fact, by including our exact results with full kinematic dependence of the coefficients of the first three leading logarithms we observe that the hadronic cross-section shifts significantly to lower values than what one obtains with the next-to-soft approximation.   

Based on these considerations, we conclude that it is not possible at this point to obtain a reliable prediction for the Higgs-boson cross-section at N$^3$LO, and that further theoretical developments are needed to achieve this goal. This is left for future work.

\section*{Acknowledgements}
The authors are grateful to Achilleas Lazopoulos and Andreas Vogt for discussions.
This research was supported by the Swiss National Science Foundation (SNF) under 
contracts 200021-143781 and  200020-149517, the European Commission 
through the ERC grants ``IterQCD'', ``LHCTheory'' (291377), ``HEPGAME'' (320651) and ``MC@NNLO'' (340983) and the FP7 Marie Curie Initial Training Network ``LHCPhenoNet'' (PITN-GA- 2010-264564), by the U.S. Department of Energy 
under contract no. DE-AC02-07CH11359 and the ``Fonds National de la Recherche Scientifique'' (FNRS), Belgium.

\end{document}